\documentclass[a4paper,11pt]{article}
\usepackage{pos}

\usepackage{mathtools}
\usepackage{bbm}
\DeclareMathAlphabet{\mathbfi}{OML}{cmm}{b}{it}

\let\originalleft\left
\let\originalright\right
\renewcommand{\left}{\mathopen{}\mathclose\bgroup\originalleft}
\renewcommand{\right}{\aftergroup\egroup\originalright}

\makeatletter
\newenvironment{equations}[1][]{\subequations\ifx\relax#1\relax\else\label{#1}\fi\align\ignorespaces}{\endalign\ignorespacesafterend\endsubequations}
\def\@spliteq#1{\begin{equation}\begin{split}#1\end{split}\end{equation}}
\def\splitequation{\collect@body\@spliteq}

\makeatother

\newcommand{\mathi}{\mathrm{i}}
\newcommand{\mathe}{\mathrm{e}}
\newcommand{\1}{\mathbbm{1}}
\newcommand{\abs}[1]{{\left\lvert{#1}\right\rvert}}
\newcommand{\sgn}{\operatorname{sgn}}
\renewcommand{\vec}[1]{{\ifnum9<1#1\mathbf{#1}\else\ifcat\noexpand#1\relax\boldsymbol{#1}\else\mathbfi{#1}\fi\fi}}
\newcommand{\bigo}[1]{\mathcal{O}\left({#1}\right)}
\newcommand{\expect}[1]{\left\langle{#1}\right\rangle}
\newcommand{\total}{\mathop{}\!\mathrm{d}}
\newcommand{\eqend}[1]{\,#1}

\title{Non-commutative coordinates from quantum gravity}

\author*[a]{Markus B. Fröb}
\author[a]{Albert Much}
\author[b]{Kyriakos Papadopoulos}

\affiliation[a]{Institut f\"ur Theoretische Physik, Universit\"at Leipzig, Br{\"u}derstra{\ss}e 16, 04103 Leipzig, Germany}

\affiliation[b]{Department of Mathematics, Kuwait University, Safat 13060, Kuwait}

\emailAdd{mfroeb@itp.uni-leipzig.de}
\emailAdd{much@itp.uni-leipzig.de}
\emailAdd{kyriakos@sci.kuniv.edu.kw}

\abstract{Local observables in (perturbative) quantum gravity are notoriously hard to define, since the gauge symmetry of gravity --- diffeomorphisms --- moves points on the manifold. In particular, this is a problem for backgrounds of high symmetry such as Minkowski space or de Sitter spacetime which describes the early inflationary phase of our universe. Only recently this obstacle has been overcome, and a field-dependent coordinate system has been constructed to all orders in perturbation theory, in which observables are fully gauge-invariant. We show that these field-dependent coordinates are non-commutative, and compute their commutator to second order in the Planck length. This provides the first systematic derivation of non-commutativity that arises due to quantum gravity effects.}

\FullConference{%
  Corfu Summer Institute 2022 ``School and Workshops on Elementary Particle Physics and Gravity'',\\
  28 August - 1 October, 2022\\
  Corfu, Greece}

\begin{document}
\maketitle

\section{Introduction}

Non-commutative geometry has long been put forward as a candidate for a theory of quantum gravity. There, the general idea that spacetime should be somehow quantised at microscopic scales has been realised in many different concrete approaches: Connes' non-commutative spectral triples~\cite{connes1995}, Lorentzian spectral triples~\cite{paschkeverch2004,franco2014}, Snyder and $\kappa$-Minkowski spacetimes and their curved-space generalisations~\cite{snyder1947,lukierskinowickiruegg1992,mignemi2010,ballesterosetal2019,franchinovinasmignemi2020}, or strict deformation quantisation~\cite{grosselechner2007,buchholzlechnersummers2011,much2012,much2017}.\footnote{Since there is a huge amount of literature on the topic, we refer to the review~\cite{douglasnekrasov2001} and the references given there and in the cited references for further works. The same caveat applies to all other references that are cited later on.} A very popular approach is the DFR spacetime~\cite{dfr1995}, where one promotes the Cartesian coordinates describing flat Minkowski spacetime to operators $\hat{x}^\mu$ and postulates commutation relations
\begin{equation}
\label{eq:xmu_xnu_theta}
[ \hat{x}^\mu, \hat{x}^\nu ] = \mathi \Theta^{\mu\nu} \eqend{,}
\end{equation}
with a constant skew-symmetric matrix $\Theta$, whose entries are supposed to be of the order of the Planck length. Such a construction solves the geometrical measurement problem~\cite{ahluwalia1993,dfr1995}, namely that the description of spacetime as a manifold must break down at distances of the order of the Planck length because any attempt to localise spacetime points with higher accurary will result in the formation of black holes and make the measurement impossible, and is thus a concrete realisation of Gedankenexperiments involving such measurements~\cite{maggiore1993,adlersantiago1999,scardigli1999}. Similar thoughts in the context of quantum mechanics of particles lead to Generalised Uncertainty Principles (GUPs)~\cite{maggiore1994,kempfmanganomann1995,scardigli1999,adlersantiago1999,scardiglicasadio2003,jizbakleinertscardigli2010,tawfikdiab2014,casadioscardigli2014,scardiglilambiasevagenas2017}, where the well-known commutator $[ x^i, p_j ] = \mathi \hbar \delta^i_j$ between position and momentum of a quantum particle is generalised to involve position and/or momentum operators on the right-hand side.

However, it seems somewhat unsatisfactory to simply postulate changed commutation relations. Instead, as the Gedankenexperiments suggest, it should be possible to derive them from gravity itself. Among the many candidates for a theory of quantum gravity, the most conservative one is perturbative Quantum Gravity (pQG), where one quantises metric fluctuations around a given classical background in the spirit of Effective Field Theory (EFT)~\cite{burgess2003}. While General Relativity (as described by the Einstein--Hilbert Lagrangian) is power-counting non-renormalisable as a quantum field theory and thus cannot be a fundamental theory, pQG is still a valid EFT where the details of the microscopic underlying theory only enter as free parameters that must be fixed by experiments. For a given accuracy of measurements, only a finite number of these parameters are relevant, and one can derive sound predictions which are valid at scales larger than the fundamental scale (which for quantum gravity is the Planck length). Since pQG results from the application of the well-established methods of Lagrangian quantum field theory to General Relativity, we expect that any theory of quantum gravity has to reproduce the predictions of perturbative quantum gravity in its region of validity, just as Newtonian gravity reproduces the results of GR in the weak gravity regime.

In pQG, the diffeomorphism invariance of General Relativity translates into a gauge symmetry for the quantised metric perturbations. Since diffeomorphisms move points and physical observables must be gauge-invariant, it is clear that local fields (i.e., defined at a point of the underlying classical spacetime manifold) cannot be observable, and the identification of observables that describe local measurements is very complicated~\cite{rovelli2000}. One possible way to construct observables that have a clear operational interpretation is the relational approach~\cite{tambornino2012}, where observables are given by the value of one dynamical field of the theory with respect to or in relation to another dynamical field, i.e., at the point where the other field has a certain value. That is, one has to measure fields not in the fixed coordinate system describing the background spacetime, but in a dynamical system (or reference frame) constructed from other fields in the theory in a suitable way. Relational observables have a long history, but some issues have only been solved recently. In particular, for highly symmetric backgrounds such as Minkowski spacetime or cosmological (Friedmann--Lemaître--Robertson--Walker) spacetimes, one cannot construct the required dynamical coordinates from curvature scalars (which have been proposed for this purpose) since these do not even distinguish all points on the background. Moreover, while it is of course possible to add dynamical scalars by hand to the theory, this changes the dynamics~\cite{gieselheroldlisingh2020}.

This problem was only solved recently~\cite{brunettietal2016,froeb2018,froeblima2018,froeblima2022}, where a dynamical coordinate system has been constructed from the gauge-dependent parts of the metric perturbation to all orders in perturbation theory. This method avoids the introduction of extra fields, and is an explicit example of the so-called geometrical clocks~\cite{giesel2008,gieselherzogsingh2018} used in the relational approach. Concretely, the dynamical coordinates are obtained as solutions of scalar differential equations in the perturbed geometry, where the choice of differential equation is related to the measurement one wants to model. The solution of these equations then depends on the metric perturbation, and since these are quantised and non-commuting, also the dynamical coordinates have a non-vanishing commutator. In this way, one obtains a non-commutative geometry from quantum gravity, which is in our opinion quite satisfactory. To make concrete predictions, we have computed this induced non-commutativity to leading order in a background Minkowski spacetime in~\cite{froebmuchpapa2023}. We explain this computation and the outcome in detail in the following sections.

\section{Observables and dynamical coordinates in perturbative quantum gravity}

As explained in the introduction, in a theory of gravity whose symmetries include diffeomorphisms the outcome of measurements can be modeled using relational observables. These are obtained by evaluating dynamical fields in a coordinate system that is also dynamical, i.e., constructed from other fields of the theory. Let us thus assume that we are given such a dynamical coordinate system $X^{(\mu)}[g,\phi]$ depending on the full dynamical metric $g_{\mu\nu}$ and possibly dynamical matter fields $\phi$. We have enclosed the index $\mu$ in parenthesis to make clear that this is not a vector, but a collection of four scalar fields which under diffeomorphisms with parameter $\xi^\mu$ transform appropriately. Let us further assume that we can separate the full metric $g_{\mu\nu}$ into a background one $g^0_{\mu\nu}$ and a perturbation $h_{\mu\nu}$:
\begin{equation}
\label{eq:relational_metricdecomp}
g_{\mu\nu} = g^0_{\mu\nu} + \kappa h_{\mu\nu} \eqend{,}
\end{equation}
where $\kappa \equiv \sqrt{16 \pi G_\text{N}}$ is a small pertubation parameter, and that the dynamical coordinates also admit a perturbative expansion:
\begin{equation}
\label{eq:relational_xmudecomp}
X^{(\mu)}(x) = x^\mu + \kappa X^{(\mu)}_1(x) + \kappa^2 X^{(\mu)}_2(x) + \bigo{\kappa^3} \eqend{.}
\end{equation}
For a background Minkowski space, we can clearly take $g^0_{\mu\nu} = \eta_{\mu\nu}$, the flat metric in Cartesian coordinates $x^\mu$, and those do discriminate between all points of the background. Perturbatively, we then also take small diffeomorphisms that keep the background fixed, $x^\mu \to x^\mu + \delta_\xi x^\mu = x^\mu - \kappa \xi^\mu$, such that the transformation of the dynamical coordinates $\delta_\xi X^{(\mu)} = \kappa \xi^\rho \partial_\rho X^{(\mu)}$ results in
\begin{equation}
\label{eq:relational_deltax1}
\delta_\xi X^{(\mu)}_1 = \xi^\mu \eqend{,} \quad \delta_\xi X^{(\mu)}_2 = \xi^\rho \partial_\rho X^{(\mu)}_1 \eqend{,}
\end{equation}
and so on. The changes in tensor fields are obtained using the Lie derivative $\mathcal{L}_{\kappa \xi}$, such that inserting the decomposition~\eqref{eq:relational_metricdecomp} with $g^0_{\mu\nu} = \eta_{\mu\nu}$ into $\delta_\xi g_{\mu\nu} = \mathcal{L}_{\kappa \xi} g_{\mu\nu}$ we obtain
\begin{equation}
\label{eq:relational_deltah}
\delta_\xi h_{\mu\nu} = \partial_\mu \xi_\nu + \partial_\nu \xi_\mu + \kappa \left( \xi^\rho \partial_\rho h_{\mu\nu} + h_{\rho\mu} \partial_\nu \xi^\rho + h_{\rho\nu} \partial_\mu \xi^\rho \right) \eqend{,}
\end{equation}
where all indices are raised and lowered using $\eta_{\mu\nu}$.

We see immediately that a first-order invariant observable can be defined by
\begin{equation}
\label{eq:relational_calh_def}
\mathcal{H}_{\mu\nu} \equiv h_{\mu\nu} - \eta_{\rho\mu} \partial_\nu X^{(\rho)}_1 - \eta_{\rho\nu} \partial_\mu X^{(\rho)}_1 + \bigo{\kappa} \eqend{,}
\end{equation}
since then the transformation~\eqref{eq:relational_deltax1} of $X^{(\mu)}_1$ cancels exactly the one of $h_{\mu\nu}$~\eqref{eq:relational_deltah}. This is of course just a special case of the general relational construction, namely evaluating the field that one is interested in (here the metric) in the dynamical coordinate system. For an arbitrary tensor field $T^{\mu \cdots}_{\nu \cdots}$, the invariant observable $\mathcal{T}^{\mu \cdots}_{\nu \cdots}$ is given by
\begin{equation}
\label{eq:relational_calt_def}
\mathcal{T}^{\mu \cdots}_{\nu \cdots}(X) \equiv \frac{\partial X^\mu}{\partial x^\alpha} \cdots \frac{\partial x^\beta}{\partial X^\nu} \cdots T^{\alpha \cdots}_{\beta \cdots}(x(X)) \eqend{,}
\end{equation}
and it is easy to check that the invariant metric perturbation~\eqref{eq:relational_calh_def} is a special case of this construction for $T_{\mu\nu} = g_{\mu\nu}$, subtracting the background metric: $\mathcal{H}_{\mu\nu} \equiv \kappa^{-1} \left( \mathcal{G}_{\mu\nu} - \eta_{\mu\nu} \right)$. That is, to obtain an invariant observable one has to invert the perturbative expansion~\eqref{eq:relational_xmudecomp} to obtain the background coordinates $x^\mu$ as functionals of the dynamical ones $X^{(\mu)}$, and then perform a change of coordinates, evaluating the field in the dynamical system (including the Jacobian factors for tensor fields). Since the dynamical coordinates transform as scalars by assumption, one can easily see that the observable $\mathcal{T}^{\mu \cdots}_{\nu \cdots}$ defined by~\eqref{eq:relational_calt_def} is invariant under diffeomorphisms, but of course this can also be checked explicitly to arbitrary order, as for $\mathcal{H}_{\mu\nu}$~\eqref{eq:relational_calh_def}.

Let us comment shortly on the philosophy behind the relational construction. What we are ultimately interested in are physical events, which can be defined by coincidences of two dynamical fields of the theory. That is, what is measured is ultimately the value of one field (the observable) at the point where another field (the coordinate, or ruler, or clock) has a given value. On the background spacetime, we can use a system of background coordinates to describe points, which we may imagine as abstraction of physical rulers and clocks. However, once gravity becomes dynamical, that is once we include a dynamical metric, we must also take into account the dynamics of the rulers and clocks themselves, and use a system of dynamical (field-dependent) coordinates to describe points. Of course, as long as we restrict to perturbation theory (such as the EFT of pQG), there is a unique map between these two sets given by the expansion~\eqref{eq:relational_xmudecomp} (and its inverse), and we can compute quantum corrections to classical results. One of these quantum effects is the non-commutativity of the coordinates.

It remains to actually construct a system of dynamical coordinates that transforms properly under diffeomorphisms and corresponds to some measurement setup. To ensure a proper transformation, one can obtain them as solutions of scalar differential equations~\cite{brunettietal2016,froeb2018,froeblima2018,lima2021} which hold on the background. For a flat Minkowski background, the Cartesian coordinates are harmonic: $\partial^2 x^\mu = 0$, and so one can define the dynamical coordinates as solutions of $\nabla^2 X^{(\mu)} = 0$~\cite{froeb2018}. Expanding metric~\eqref{eq:relational_metricdecomp} and dynamical coordinates~\eqref{eq:relational_xmudecomp} to first order, for the first-order correction we obtain~\cite{froebmuchpapa2023}
\begin{equation}
\label{eq:relational_x1sol}
X^{(\mu)}_1(x) = \int G(x,y) H^\mu(y) \total^4 y \eqend{,} \quad H^\mu \equiv \partial_\rho h^{\rho\mu} - \frac{1}{2} \partial^\mu h \eqend{,}
\end{equation}
where $G$ is a Green's function of the flat d'Alembertian: $\partial^2 G(x,y) = \delta^4(x-y)$. Clearly, the $X^{(\mu)}_1$ are non-local functionals of the metric perturbation $h_{\mu\nu}$, and thus the invariant observables~\eqref{eq:relational_calt_def} will be as well, as explained in the introduction. However, by a proper choice of $G$ we can at least obtain a causal observable and avoid unphysical action-at-a-distance effects. In the classical theory, this forces us to take the retarded Green's function $G_\text{ret}(x,y) = \Theta(x^0-y^0) \delta[(x-y)^2]$ which has support on the past lightcone; in the quantum theory, we have to use the in-in formalism~\cite{chousuhaolu1985,jordan1986} to ensure a causal evolution. Since the background Cartesian coordinates determine straight lines, the dynamical coordinates $X^{(\mu)}$ defined as solutions of $\nabla^2 X^{(\mu)} = 0$ in a sense determine lines which are as straight as possible in the perturbed geometry. In the next section, we will see that they indeed do not commute anymore.

\section{Non-commutative coordinates}

The basic reason why the dynamical coordinates do not commute is that they are functionals of the dynamical fields, which have a non-vanishing commutator after quantisation. Therefore, the non-commutativity of the coordinates is not fundamental, but induced from the one of the fields. In pQG, we are quantising the metric perturbation $h_{\mu\nu}$~\eqref{eq:relational_metricdecomp}, whose action is obtained by expanding the Einstein--Hilbert action for gravity $S_\text{EH} = \kappa^{-2} \int R \sqrt{-g} \total^4 x$ to second order around the flat Minkowski background and adding the standard de Donder gauge term~\cite{capperleibbrandtramonmedrano1973} to fix the gauge transformation~\eqref{eq:relational_deltah}. By neglecting all higher-order terms coming from the expansion of the Einstein--Hilbert action, we are treating $h_{\mu\nu}$ as a free quantum field on the Minkowski background, which gives the leading EFT corrections, in particular the leading-order commutator of the dynamical coordinates. For a free theory, Wick's theorem determines all correlation functions in terms of the (time-ordered) two-point function
\begin{equation}
\label{eq:coords_feynman}
G^\text{F}_{\mu\nu\rho\sigma}(x,x') \equiv - \mathi \expect{ \mathcal{T} h_{\mu\nu}(x) h_{\rho\sigma}(x') } = \left( 2 \eta_{\mu(\rho} \eta_{\sigma)\nu} - \eta_{\mu\nu} \eta_{\rho\sigma} \right) G^\text{F}(x,x') \eqend{,}
\end{equation}
where
\begin{equation}
\label{eq:coords_feynman_scalar}
G^\text{F}(x,x') = \int \tilde{G}^\text{F}(\vec{p},t,t') \, \mathe^{\mathi \vec{p} (\vec{x}-\vec{x}')} \frac{\total^3 \vec{p}}{(2\pi)^3} \eqend{,} \quad \tilde{G}^\text{F}(\vec{p},t,t') \equiv - \mathi \frac{\mathe^{- \mathi \abs{\vec{p}} \abs{t-t'}}}{2 \abs{\vec{p}}}
\end{equation}
is the massless scalar time-ordered two-point function or Feynman propagator. Moreover, for a free theory the commutator of two $X^{(\mu)}_1$ will be proportional to the identity since they are linear functionals of $h_{\mu\nu}$~\eqref{eq:relational_x1sol}, and we can thus obtain the commutator by computing its expectation value:
\begin{equation}
\label{eq:coords_commutator}
\left[ X^{(\mu)}_1(x), X^{(\nu)}_1(x') \right] = \expect{ \left[ X^{(\mu)}_1(x), X^{(\nu)}_1(x') \right] } \1 \eqend{.}
\end{equation}

To compute a true expectation value instead of an S-matrix element, we need to use the in-in or closed-time-path formalism~\cite{chousuhaolu1985,jordan1986}, which also ensures a causal evolution. In the in-in formalism, the time integration is replaced by the integration along a path going from $-\infty$ to $+\infty$ and back to $-\infty$, and the time-ordering of fields in the usual in-out formalism is replaced by the path-ordering of fields along this integration contour. Depending on which part of the contour the fields are put, we thus have the corresponding path-ordered two-point functions
\begin{equation}
\label{eq:coords_gab_munurhosigma}
G^{AB}_{\mu\nu\rho\sigma}(x,x') \equiv - \mathi \expect{ \mathcal{P} h^A_{\mu\nu}(x) h^B_{\rho\sigma}(x') }
\end{equation}
with $A,B = \pm$, which have the same form as~\eqref{eq:coords_feynman}, but with the scalar Feynman propagator~\eqref{eq:coords_feynman_scalar} replaced by the corresponding Wightman function or anti-time-ordered Dyson propagator:
\begin{equations}
G^{++}_{\mu\nu\rho\sigma}(x,x') &= - \mathi \expect{ \mathcal{T} h^+_{\mu\nu}(x) h^+_{\rho\sigma}(x') } = \left( 2 \eta_{\mu(\rho} \eta_{\sigma)\nu} - \eta_{\mu\nu} \eta_{\rho\sigma} \right) G^\text{F}(x,x') \eqend{,} \\
G^{+-}_{\mu\nu\rho\sigma}(x,x') &= - \mathi \expect{ h^-_{\rho\sigma}(x') h^+_{\mu\nu}(x) } = \left( 2 \eta_{\mu(\rho} \eta_{\sigma)\nu} - \eta_{\mu\nu} \eta_{\rho\sigma} \right) G^-(x,x') \eqend{,} \\
G^{-+}_{\mu\nu\rho\sigma}(x,x') &= - \mathi \expect{ h^-_{\mu\nu}(x) h^+_{\rho\sigma}(x') } = \left( 2 \eta_{\mu(\rho} \eta_{\sigma)\nu} - \eta_{\mu\nu} \eta_{\rho\sigma} \right) G^+(x,x') \eqend{,} \\
G^{--}_{\mu\nu\rho\sigma}(x,x') &= - \mathi \expect{ \overline{\mathcal{T}} h^-_{\mu\nu}(x) h^-_{\rho\sigma}(x') } = \left( 2 \eta_{\mu(\rho} \eta_{\sigma)\nu} - \eta_{\mu\nu} \eta_{\rho\sigma} \right) G^\text{F}(x,x') \eqend{,}
\end{equations}
since fields on the backward part of the contour are always ``later'' than fields on the forward part. The scalar propagators in Fourier space read
\begin{equation}
\label{eq:coords_tildeg_scalar}
\tilde{G}^+(\vec{p},t,t') \equiv - \mathi \frac{\mathe^{- \mathi \abs{\vec{p}} (t-t')}}{2 \abs{\vec{p}}} \eqend{,} \quad \tilde{G}^-(\vec{p},t,t') \equiv - \mathi \frac{\mathe^{\mathi \abs{\vec{p}} (t-t')}}{2 \abs{\vec{p}}} \eqend{,} \quad \tilde{G}^\text{D}(\vec{p},t,t') \equiv - \mathi \frac{\mathe^{\mathi \abs{\vec{p}} \abs{t-t'}}}{2 \abs{\vec{p}}}
\end{equation}
and we see that they only differ from the Feynman propagator in the time dependence in the exponential. The expectation value of the commutator~\eqref{eq:coords_commutator} can thus be written as
\begin{equation}
\label{eq:coords_commutator_expect}
\expect{ \left[ X^{(\mu)}_1(x), X^{(\nu)}_1(x') \right] } = \expect{ \mathcal{P} X^{-(\mu)}_1(x) X^{+(\nu)}_1(x') } - \expect{ \mathcal{P} X^{+(\mu)}_1(x) X^{- (\nu)}_1(x') }
\end{equation}
with
\begin{equation}
\label{eq:coords_expect}
\expect{ \mathcal{P} X^{A(\mu)}_1(x) X^{B(\nu)}_1(x') } = \iint G^{AC}(x,y) G^{BD}(x',y') H^\mu_C(y) H^\nu_D(y') \total^4 y \total^4 y' \eqend{,}
\end{equation}
where we used the explicit expression~\eqref{eq:relational_x1sol} for the first-order dynamical coordinates $X^{(\mu)}_1$, and where the Einstein summation convention also holds for $C,D = \pm$. Note that all time integrals in the in-in formalism are taken over the closed contour, which includes the integrals in~\eqref{eq:relational_x1sol}; the required propagators are just the scalar ones~\eqref{eq:coords_feynman_scalar} and~\eqref{eq:coords_tildeg_scalar}. In the classical limit, one does not distinguish between the ``$+$'' and ``$-$'' fields, and then in the expectation value~\eqref{eq:coords_expect} one encounters either the combinations $G^\text{F} - G^-$ (for $A/B = +$) or $G^+ - G^\text{D}$ (for $A/B = -$), which are both equal to the retarded propagator $G^\text{ret}$. In this way the in-in formalism ensures a causal evolution~\cite{chousuhaolu1985,jordan1986}.

To finally compute the expectation value~\eqref{eq:coords_expect}, we also need to ensure that we choose the right state. Since the massless propagators~\eqref{eq:coords_feynman_scalar} and~\eqref{eq:coords_tildeg_scalar} only decay like a power for large separations, the integrals harbor potential IR divergences coming from the integration over the past lightcone. To obtain an IR-finite result, we need to slightly deform the integration contour in the complex plane, which is the correct prescription to select the true interacting vacuum state of the theory~\cite{peskinschroeder,froebrouraverdaguer2012,baumgartsundrum2021}. In practice, this amounts to adding a convergence factor $\exp[\epsilon \abs{\vec{p}} (y^0+y'^0)]$ with $\epsilon > 0$ to the integrals~\eqref{eq:coords_expect}, and take the limit $\epsilon \to 0$ after integration. Even though we consider a theory of free fields, there are non-trivial interactions because of the integration over the past light cone; the convergence factor can thus also be interpreted as an adiabatic cutoff of the interaction in the far past~\cite{lippmannschwinger1950}. Using the expressions of the propagators in Fourier space~\eqref{eq:coords_feynman_scalar} and~\eqref{eq:coords_tildeg_scalar}, the integrals in~\eqref{eq:coords_expect} including the convergence factors can then be easily computed, and we obtain
\begin{equation}
\label{eq:coords_commutator_expect_fourier}
\expect{ \left[ X_1^{(\mu)}(x), X_1^{(\nu)}(x') \right] } = \int \frac{\mathi \eta^{\mu\nu}}{2 \abs{\vec{p}}^3} \Big[ \cos\left[ \abs{\vec{p}} (t-t') \right] \abs{\vec{p}} (t-t') - \sin\left[ \abs{\vec{p}} (t-t') \right] \Big] \mathe^{\mathi \vec{p} (\vec{x}-\vec{x}')} \frac{\total^3 \vec{p}}{(2\pi)^3} \eqend{.}
\end{equation}
To transform this expression back into real space, we use spherical coordinates, and a lengthy but ultimately straightforward computation leads to~\cite{froebmuchpapa2023}
\begin{splitequation}
\label{eq:coords_commutator_result}
\left[ X_1^{(\mu)}(x), X_1^{(\nu)}(x') \right] = - \mathi \frac{\eta^{\mu\nu}}{8 \pi} \sgn(t-t') \Theta[ -(x-x')^2 ] \1 \eqend{.}
\end{splitequation}
We see that the commutator of two $X_1^{(\mu)}$ vanishes for spacelike separations (due to the Heaviside $\Theta$ function). For timelike separations, we obtain a constant with the sign depending on whether the second $X_1^{(\nu)}$ is in the future or in the past of the first $X_1^{(\mu)}$. Moreover, our result is fully Lorentz-invariant.

Finally we can compute the commutator of two dynamical coordinates $X^{(\mu)}$ to leading order. For this, we use that the background coordinates $x^\mu$ commute with everything, such that
\begin{splitequation}
\label{eq:coords_commutator_fullresult}
\left[ X^{(\mu)}, Y^{(\nu)} \right] &= \kappa^2 \left[ X_1^{(\mu)}(x), X_1^{(\nu)}(y) \right] + \bigo{\kappa^3} \\
&= - \mathi \kappa^2 \frac{\eta^{\mu\nu}}{8 \pi} \sgn(X^0-Y^0) \Theta[ -(X-Y)^2 ] + \bigo{\kappa^3} \eqend{.}
\end{splitequation}
Using that $\kappa^2 = 16 \pi G_\text{N} = 16 \pi \ell_\text{Pl}^2$, we see that the leading-order result is proportional to the squared Planck length $\ell_\text{Pl}$, which appears naturally. The commutator~\eqref{eq:coords_commutator_fullresult} is the main result of our work~\cite{froebmuchpapa2023}.

\section{Discussion and outlook}

Comparing our result~\eqref{eq:coords_commutator_fullresult} with the commutation relations~\eqref{eq:xmu_xnu_theta} of the DFR spacetime, there are two main differences: first, the commutator is not constant, but depends on the coordinates themselves. Clearly, the most general form of the commutator reads
\begin{equation}
[ X^\mu, Y^\nu ] = \mathi \Theta^{\mu\nu}(X,Y) \eqend{,}
\end{equation}
with the matrix $\Theta$ fulfilling
\begin{equation}
\Theta^{\mu\nu}(X,Y) = - \Theta^{\nu\mu}(Y,X) \eqend{,} \quad [ \Theta^{\mu\nu}(X,Y) ]^\dagger = \Theta^{\mu\nu}(X,Y)
\end{equation}
to ensure antisymmetry and reality of the commutator, the second because the $X^\mu$ are Hermitean operators. In the DFR spacetime, this is solved by taking $\Theta$ constant, real, and antisymmetric in its indices. For our result, the matrix $\Theta^{\mu\nu}(X,Y) = - 2 \ell_\text{Pl}^2 \eta^{\mu\nu} \sgn(X^0-Y^0) \Theta[ -(X-Y)^2 ]$ depends on the coordinates $X$ and $Y$, and so can be symmetric in its indices since the antisymmetry comes from the sign $\sgn(X^0-Y^0)$. The second (and more important) difference concerns the interpretation of the coordinate operators $X^{(\mu)}$. In the DFR approach (and others such as spectral triples), the non-commutative coordinates are operators in some abstract space, and the classical geometry only emerges from their spectrum: the physical coordinates are the (possibly generalized) eigenvalues of the coordinate operators. In pQG, we instead associate a dynamical field-dependent coordinate to each physical event. That is, once one takes dynamical gravity into account, one cannot work anymore with a fixed background, but has to include the dynamics of the reference frame itself. Physical events are thus described in a relational way, by the state of one dynamical field of the theory with respect to this dynamical coordinate system. Since the dynamical coordinates are constructed from the metric perturbation~\eqref{eq:relational_x1sol}, they become non-commuting operators in the quantum theory. In this way, quantum gravity induces a non-commutativity of the coordinates that are needed to describe observables in the full theory.

Using the well-known formula
\begin{equation}
\Delta_A \Delta_B \geq \frac{1}{2} \abs{ \expect{ [A, B] } }
\end{equation}
relating the standard deviations $\Delta$ of two Hermitean operators $A$ and $B$ (that is, the uncertainties in the measurement of their values) to the expectation value of their commutator, we obtain from the result~\eqref{eq:coords_commutator_fullresult} a generalised uncertainty principle
\begin{equation}
\label{eq:nc_gup}
\Delta_X \Delta_Y \geq \ell_\text{Pl}^2 \, \Theta[ -(X-Y)^2 ] \eqend{.}
\end{equation}
Taken at face value, this uncertainty principle tells us that the measurements of coordinates whose corresponding events are timelike separated (for which $\Theta[ -(X-Y)^2 ] = 1$) are always uncertain, with a standard deviation of exactly the Planck length (up to higher-order corrections). On the other hand, measurements of coordinates whose corresponding physical events have spacelike separation (for which $\Theta[ -(X-Y)^2 ] = 0$) are uncorrelated and certain. However, in contrast to the well-known quantum-mechanical Heisenberg uncertainty principle relating the uncertainties in the measurement of a particle's position and momentum, in quantum gravity it is impossible to repeatedly measure the coordinates $X^{(\mu)}$ of the \emph{same} event. The operational meaning of the standard deviation $\Delta_X$ is therefore not clear, and we leave the question how one can extract observable results from the commutator~\eqref{eq:coords_commutator_fullresult} to future work. What we can assert, and what is reflected in the vanishing of the commutator~\eqref{eq:coords_commutator_fullresult} for spacelike separations, is that a measurement can not influence other measurements at spacelike separations. This statement of the microcausality principle, namely the impossibility of superluminal signaling, holds independently of the state in which the system is prepared~\cite{hellwigkraus1970,doplicher2018,fewsterverch2018}. In our case, it is ultimately a consequence of the causal evolution of the field-dependent coordinates guaranteed by the in-in formalism, and the fact that the gravitational perturbation $h_{\mu\nu}$ satisfies microcausality since it is quantized according to the standard rules for relativistic quantum fields.

As for any EFT, our result~\eqref{eq:coords_commutator_fullresult} is valid at length scales above the Planck length, where one can neglect higher orders in comparison to the leading one. In particular, we note that the causal relation between the events described by the non-commutative coordinates is to leading order the same as the one of the commuting background coordinates, and any non-commutativity would only appear at higher orders; the right-hand side of~\eqref{eq:coords_commutator_fullresult} is thus unambiguous. For the same reason, we cannot infer strong statements about the resolution of singularities from our result (as is done for black holes~\cite{nicolinismailagicspalluci2006}), but of course our result gives hints on how a fundamental quantum gravity theory could naturally incorporate non-commutativity. To obtain a result that is also valid for smaller distances, we would have to compute higher-order corrections. An important question that then arises is the proper interpretation of a causal relation between non-commuting coordinates, and the topology of the manifold described by these coordinates. While there are some proposals (for example for Riemannian spectral triples~\cite{connes2013}), this is a challenge especially in the Lorentzian case~\cite{besnard2009}. Already in the classical case, there can be a mismatch between the topology of the underlying manifold and the causal ordering induced by the Lorentzian metric~\cite{papadopoulos2021}; see~\cite{finstermuchpapadopoulos2021} for an overview of results and open questions.

Nevertheless, before investigating in detail these foundational questions one needs to collect more information in concrete settings. We are currently generalising our results to a background de Sitter spacetime which describes the current accelerated expansion of our universe; we plan furthermore to study non-commutative coordinates for pQG around a cosmological (FLRW) background which is relevant during inflation, the period of primordial expansion. An important question is the choice of dynamical coordinate system. In principle, it should be chosen in accordance with the experimental setup, or otherwise said, we need to model which rulers and clocks are actually used to measure observables. The generalized harmonic coordinates $\nabla^2 X^{(\mu)} = 0$ that we used here are a natural choice that also appears in other contexts, for example matrix models~\cite{steinacker2010}, but one might also contemplate other choices such as geodesic lightcone coordinates~\cite{prestonpoisson2006,gasperinimarozzinugierveneziano2011,fanizzamarozzimedeirosschiaffino2021,mitsoufanizzagrimmyoo2021,froeblima2022} which model observations made along the observer's past lightcone, or synchronous coordinates~\cite{froeblima2023}. Last but not least, it remains to derive observational consequences of the non-commutativity~\eqref{eq:coords_commutator_fullresult} and its future generalizations to other backgrounds.

\acknowledgments

M.B.F. acknowledges the support by the Deutsche Forschungsgemeinschaft (DFG, German Research Foundation) --- project no. 396692871 within the Emmy Noether grant CA1850/1-1 and project no. 406116891 within the Research Training Group RTG 2522/1. A.M. acknowledges the support by the DFG within the Sonderforschungsbereich (SPP, Priority Program) 2026 ``Geometry at Infinity''. We thank Rainer Verch for comments, Roberto Casadio and Fabio Scardigli for reference suggestions, and the participants of the 2022 Corfu ``Workshop on Noncommutative and generalized geometry in string theory, gauge theory and related physical models'', in particular Paolo Aschieri, Eugenia Boffo and Harold Steinacker, as well as Nikolaos Kalogeropoulos for discussions, comments and reference suggestions.

\bibliographystyle{JHEP}
\setlength{\bibsep}{4.9pt plus 0.3ex}
\bibliography{literature}

\end{document}